\documentstyle[aps,prl]{revtex}

\def\be{\begin{equation}}
\def\ee{\end{equation}}
\def\bea{\begin{eqnarray}}
\def\eea{\end{eqnarray}}
\def\bea*{\begin{eqnarray*}}
\def\eea*{\end{eqnarray*}}
\def\bit{\begin{itemize}}
\def\eit{\end{itemize}}

\begin{document}
\title{One and two-component hard-core plasmas}

\author{
Roland R. Netz$^{\$}$ and 
Henri Orland$^*$
}
\address{$^{\$}$Max-Planck-Institut f\"ur Kolloid- 
und Grenzfl\"achenforschung,
Kantstr. 55, 14513 Teltow, Germany}
\address{$^*$Service de Physique Th\'eorique, 
CEA-Saclay, 91191 Gif sur Yvette, France}

\date{\today}
\maketitle

\begin{abstract}
A field theory is presented for particles
which interact via Coulomb and hard-core potentials.
We apply the method to the one-component plasma 
(OCP) with hard cores, consisting of 
identical particles of fixed charge and diameter in
a neutralizing background, and the symmetric two-component
plasma (TCP) with hard cores, 
consisting of equal numbers of positively and negatively
charged particles of identical size. We obtain exactly the 
first few coefficients of  a systematic  low-density
expansion of the free energy for both models. 
The OCP coefficients  go over to
the classical Abe result in the high-charge 
(or small hard-core diameter) limit.
The TCP, on the
other hand, exhibits diverging coefficients in the 
high-charge limit, which is due to the formation of
strongly bound  ion-pairs. 
\end{abstract}

\bigskip
\noindent PACS. 61.25.Hq, 83.70.Hq, 61.41.+e \\

\maketitle

\section{Introduction}

The equation of state of a one-component plasma (OCP), i.e. of a system
consisting of point-like charged particles, is of central importance
for plasma theory\cite{rev1}. It is commonly assumed that the 
OCP follows from the two-component plasma (TCP), consisting
of a globally neutral mixture of positive and negative point particles,
if one charge species is of lesser charge
and forms a neutralizing background.
Much interest in the plasma community
has been focused on the limit of high  plasma density (or large plasma
parameter), which defines the limit when the electrostatic interaction
at average separation between two particles is much larger than thermal
energy\cite{rev2}. This situation is, for example, realized in white dwarfs and
the interior of large planets (such as Jupiter).
Our interest in plasmas comes mainly from the close analogy with 
colloidal and electrolyte solutions, which also consist of charged
particles. Very often in colloidal systems,
the concentration of charged
species is rather low, such that a low-density expansion becomes 
meaningful. Also, in addition to Coulombic interactions, the excluded-volume
interaction becomes important in colloidal and electrolyte systems,
and therefore has to be included\cite{Borukhov}.
The OCP without hard core interactions has been studied by a variety of methods\cite{rev1}.
The first few terms in a low-density expansion of the free energy have been 
obtained by an explicit resummation of the Mayer expansion\cite{Abe}.
The OCP with hard cores has in the past been studied extensively with
numerical and integral-equation methods\cite{OCPhardcore}.
For the TCP, the leading term in the free energy has been calculated
by Debye and H\"uckel in their classical paper\cite{Debye}.
The formal foundation for the rigorous thermodynamic treatment  
of a TCP including a hard-core interaction has been given by 
Mayer\cite{Mayer} and Edwards\cite{Edwards},
 but to our knowledge,
no systematic calculation of the low density expansion of the free
energy has been performed.
 
In this article,
we present a novel field-theoretic method which can be used 
to systematically calculate  free  energies and other quantities
of interest for systems of charged hard-core particles
in powers of particle densities. We apply this method both to 
the OCP and TCP in the presence of finite hard-core radii. 
In the limit of vanishing particle charges,
we recover the virial expansion of a hard core gas for both the OCP and TCP.
In the limit of vanishing hard-core radius, we recover the classical
Abe result for the OCP\cite{Abe}. The expansion coefficients of the TCP 
diverges in the limit of vanishing hard-core radius, due
to the proliferation of bound particle pairs. 
Much interest has been focused on the critical behavior of the TCP
with hard core interactions(the so-called restricted
primitive model, RPM)\cite{Stell,Fisher,Sengers,Brilliantov,Netz}.
Our results demonstrate that the critical
point of the RPM is not accessible by a low-density expansion.

In the next two sections we present our methods and calculations
for the TCP and OCP with hard cores, respectively. Our main results consist of
the density expansion coefficients of the free energy, which are obtained
in closed form for the full range of coupling constants. 
The final section is devoted to a brief discussion of our results.

\section{Two-component plasma}

We consider a gas of $N_+ $ positive charges and $N_- $ negative charges,
interacting with Coulomb interactions 
$v({\bf r}) = \ell_B/ r$ and an additional short-ranged potential 
$w({\bf r}) $ which we assume to be a hard-core interaction
of range $a$. The Bjerrum length $\ell_B = q^2 /4 \pi \varepsilon k_B T$
measures the distance at which the interaction between
two charged particles equals the thermal energy.
The canonical partition function of the system reads
\begin{eqnarray}
Z_{N_+,N_-}  &=& \frac{1}{N_+! N_-! } \prod_{i=1}^{N_+} \left[
\int {\rm d} {\bf r}_i^+ \right]  \prod_{i=1}^{N_-} \left[ 
\int {\rm d} {\bf r}_i^- \right]
\exp \left\{ -\frac 1 2 \int {\rm d} {\bf r}  {\rm d} {\bf r}'
\left[ \rho ({\bf r})w({\bf r-r}')\rho ({\bf r}')+
  \rho_c({ \bf r})v({\bf r-r}')\rho _c({\bf r}') \right] \right\}
  \nonumber \\ && {\rm e}^{(N_+ +N_-) [w(0)+v(0)]/2} 
\label{partition}
\end{eqnarray}
where \[ \rho ({\bf r})=\sum_{i=1}^{N_+}\delta 
({\bf r}-{\bf r}_i^+)+\sum_{i=1}^{N_-}
\delta ({\bf r}-{\bf r}_i^-) \]
denotes the particle density and 
\[ \rho _c({\bf r})=\sum_{i=1}^{N_+}\delta
({\bf r}-{\bf r}_i^+)-\sum_{i=1}^{N_-} \delta ({\bf r}-{\bf r}_i^-)\]
  is the charge density.
The infinite particle self-energies $v(0)$ and $w(0)$ are subtracted.
We perform a double Stratanovitch-Hubbard transform and get
\begin{eqnarray}
Z_{N_+,N_-} &=& \int \frac{{\cal D} \phi}{Z_v} 
\int \frac{{\cal D} \psi}{Z_w} 
\exp \left\{ -\frac 1 2 \int {\rm d} {\bf r}  {\rm d} {\bf r}'
\left[ \psi ({\bf r})w^{-1} ({\bf r-r}')\psi ({\bf r}')+
  \phi({\bf r})v^{-1}({\bf r-r}')\phi({\bf r}') \right] \right\}
  \nonumber \\ && 
\frac{1}{N_+!} \left[ \int {\rm d} {\bf r}  {\rm e}^{-i \psi({\bf r}) 
-i \phi({\bf r}) +w(0)/2+v(0)/2} \right]^{N_+}
\frac{1}{N_-!} \left[ \int {\rm d} {\bf r}  {\rm e}^{-i \psi({\bf r}) 
+i \phi({\bf r}) +w(0)/2+v(0)/2} \right]^{N_-}
\end{eqnarray}
where $Z_v$ and $Z_w$ denote the square root of the 
determinants of the Coulomb and
hard-core operators, $Z_v \sim \sqrt{\det v}$ and $Z_w \sim \sqrt{\det w}$,
respectively.
Assuming that the fugacities of the
positive and negative ions are the same,
the grand-canonical partition function is defined by
\begin{equation}
\label{granddef}
Z_\lambda = \sum_{N_+,N_-} \lambda^{N_+ + N_-} Z_{N_+,N_-} 
\end{equation}
and reads 
\begin{eqnarray}
Z_\lambda &=& \int \frac{{\cal D} \phi}{Z_v} 
\int \frac{{\cal D} \psi}{Z_w} 
\exp \left\{ -\frac 1 2 \int {\rm d} {\bf r}  {\rm d} {\bf r}'
\left[ \psi ({\bf r})w^{-1} ({\bf r-r}')\psi ({\bf r}')+
  \phi({\bf r})v^{-1}({\bf r-r}')\phi({\bf r}') \right] \right.
  \nonumber \\ && \left.
+ 2 \lambda  \int {\rm d} {\bf r} h({\bf r})
{\rm e}^{ v(0)/2} \cos[\phi({\bf r})]
 \right\}
\end{eqnarray}
where we used the definition
\begin{equation} \label{hdef}
h({\bf r}) \equiv {\rm e}^{-i \psi({\bf r}) 
 +w(0)/2}.
 \end{equation}
We next introduce the Debye-H\"uckel propagator, whose inverse
is given by the ordinary screened Coulomb potential
$v_{\rm DH}({\bf r})= \ell_B {\rm e}^{-\kappa r}/r$
with $\kappa^2 = 8 \pi \ell_B \lambda$. The
partition function can be written as
\begin{equation} 
Z_\lambda ={\rm e}^{2\lambda V+\lambda V v(0)}
\frac{Z_{\rm DH}}{Z_v} \left\langle 
{\rm e}^{ 2 \lambda  \int {\rm d} {\bf r} Q({\bf r})
  }\right\rangle
 \end{equation}
where 
\begin{equation}
Q({\bf r}) \equiv 
h({\bf r}) {\rm e}^{ v(0)/2} \cos[\phi({\bf r})]
-1-v(0)/2+ \phi^2({\bf r})/2
 \end{equation}
and the average brackets denote averages with respect to
the DH propagator and the hard-core propagator. 
The determinant of the Debye-H\"uckel interaction is denoted
as $Z_{\rm DH} \sim \sqrt{\det v_{\rm DH}}$ and will be explicitly 
calculated below.
Defining the
effective interaction $Q$  in this fashion allows us to perform a systematic
low-density expansion, since all higher cumulant expectation values of 
$Q$ are finite even if the fugacity inside the Debye-H\"uckel propagator
is set to zero.
The logarithm of the 
partition function is given by a cumulant expansion in powers
of $Q$,
\begin{equation} \label{logZ}
\frac{\ln Z_\lambda}{V} = 2\lambda +\lambda  v (0) + V^{-1} \ln \left[
\frac{Z_{\rm DH}}{Z_v} \right] +2 \lambda Z_1 +
2 \lambda^2 Z_2 + \cdots 
\end{equation}
where
\begin{equation} \label{Z1def}
Z_1 = V^{-1} \int {\rm d}{\bf r} \langle Q ({\bf r}) \rangle,
\end{equation}
\begin{equation} \label{Z2def}
Z_2 = V^{-1} \int {\rm d}{\bf r}_1 {\rm d}{\bf r}_2 
\langle Q({\bf r}_1)  Q({\bf r}_2)  \rangle-  V^{-1} \left[
\int {\rm d}{\bf r} \langle Q ({\bf r}) \rangle  \right]^2 .
\end{equation}
All  expectation values appearing in these expressions
can be calculated  explicitly. We note that, neglecting
the hard-core interaction,  the cumulant 
terms $Z_n$ correspond to  giant cluster graphs considered
by Abe\cite{Abe}. 
Since the integrals with respect
 to the fluctuating $\phi$ and $\psi$ fields decouple, 
we do not  have to 
consider  mixed expectation values. Up to second order in $Q$ we
only need the expectation values
\begin{equation}
 \langle h({\bf r}) \rangle=1,
\end{equation}
\begin{equation}
 \langle h({\bf r}_1) h({\bf r}_2 ) \rangle={\rm e}^{-w({\bf r}_1-{\bf 
 r}_2 )},
\end{equation}
\begin{equation}
 \langle \phi^2({\bf r}) \rangle=v_{\rm DH}(0),
\end{equation}
\begin{equation}
 \langle \phi^2({\bf r}_1) \phi^2({\bf r}_2) \rangle=
2 v^2_{\rm DH} ({\bf r}_1-{\bf r}_2 )+ v^2_{\rm DH}(0),
\end{equation}
\begin{equation}
 \langle \cos[\phi({\bf r})] \rangle={\rm e}^{-v_{\rm DH}(0)/2},
\end{equation}
\begin{equation}
 \langle \cos[\phi({\bf r}_1)] \cos[\phi({\bf r}_2)] \rangle=
 {\rm e}^{-v_{\rm DH}(0)}\left({\rm e}^{-v_{\rm DH}({\bf r}_1-{\bf r}_2)}
 + {\rm e}^{v_{\rm DH}({\bf r}_1-{\bf r}_2)} \right)/2,
\end{equation}
\begin{equation}
 \langle \phi^2({\bf r}_1) \cos[\phi({\bf r}_2)] \rangle=
 {\rm e}^{-v_{\rm DH}(0)/2}\left(v_{\rm DH}(0)
-  v^2_{\rm DH}({\bf r}_1-{\bf r}_2) \right).
\end{equation}
Introducing the difference between the Debye-H\"uckel and 
the bare Coulomb self energy, which is finite, 
\begin{equation}
\Delta v_0 = v(0)-v_{\rm DH}(0)= \ell_B \sqrt{8 \pi \ell_B \lambda},
\end{equation}
the functions $Z_1$ and $Z_2$ can be explicitly written as 
\begin{equation}
\label{Z1res}
Z_1 ={\rm e}^{\Delta v_0/2} -1 - \Delta v_0/2,
\end{equation}
\begin{equation}
\label{Z2res}
Z_2 = \int {\rm d}{\bf r} \left[\frac{1}{2} 
{\rm e}^{\Delta v_0-w({\bf r})} \left(
{\rm e}^{-v_{\rm DH}({\bf r})}
 + {\rm e}^{v_{\rm DH}({\bf r})} \right)
 -{\rm e}^{\Delta v_0}+\frac{1}{2} v^2_{\rm DH}({\bf r})
 \left(1-2 {\rm e}^{\Delta v_0/2}\right) \right].
\end{equation}
The DH contribution to the free energy in eq.(\ref{logZ}) follows as
\begin{equation}
\label{DHres}
\lambda  v(0) + V^{-1} \ln \left[
\frac{Z_{\rm DH}}{Z_v} \right] =
-\frac{1}{2} \int \frac{{\rm d} {\bf q}}{(2 \pi)^3}
\ln\left(1+\frac{8 \pi \ell_B \lambda}{q^2} \right)
+ \lambda \int \frac{{\rm d} {\bf q}}{(2 \pi)^3}
\frac{4\pi \ell_B}{q^2}
=\frac{(8 \pi \ell_B \lambda)^{3/2}}{12 \pi}.
\end{equation}
Next we have to systematically carry out the back-Legendre
transform to the canonical ensemble. 
To do this, we expand the grand-canonical potential 
in eq.(\ref{logZ}) in 
fractional powers of the rescaled fugacity, $\tilde{\lambda} 
= a^3 \lambda$,
\begin{equation} \label{expand}
\frac{a^3 \ln Z_\lambda}{V} = 2 \tilde{\lambda}
 - b_{3/2} \tilde{\lambda}^{3/2} - b_2
\tilde{\lambda}^2  -b_{5/2} \tilde{\lambda}^{5/2} - \cdots
\end{equation}
From the definition of the grand-canonical partition function,
eq.(\ref{granddef}), it follows  
that the the rescaled concentration of ion pairs,
$\tilde{c} = a^3(N_++N_-)/2V$, is given by
\begin{equation} \label{ctilde}
\tilde{c} = \frac{ \lambda \partial a^3 \ln Z_\lambda /V }
{  2 \partial \lambda}.
\end{equation}
From eqs.(\ref{expand}) and (\ref{ctilde})
we obtain the fugacity as a function of the concentration
\begin{equation}
\tilde{\lambda}= \tilde{c}  +\frac{3}{4} b_{3/2} \tilde{c}^{3/2}  +
\left(b_2+\frac{27}{32} b_{3/2}^2 \right) \tilde{c}^2 + 
\left(\frac{5}{4} b_{5/2}+\frac{21}{8} b_2 b_{3/2}
+ \frac{567}{512} b_{3/2}^3 \right) \tilde{c}^{5/2}
+\cdots
\end{equation}
Inserting this into eq.(\ref{expand}),
the rescaled free energy density follows as 
\begin{eqnarray}
a^3 f &=& - \frac{ a^3 \ln Z_\lambda}{V} + 
2 \tilde{c} \ln \tilde{\lambda} \nonumber \\
&=& 2\tilde{c} \ln \tilde{c} -2\tilde{c} +d_{3/2} \tilde{c}^{3/2} +
d_2  \tilde{c}^2 + d_{5/2} \tilde{c}^{5/2} +\cdots
\end{eqnarray}
with the expansion coefficients given by 
\[d_{3/2} = b_{3/2}, \]
\[ d_2 = b_2 + \frac{9}{16} b_{3/2}^2, \]
\[ d_{5/2} = b_{5/2}+ \frac{3}{2} b_2 b_{3/2} + \frac{63}{128} b_{3/2}^3.\]
To carry out the expansion of the grand-canonical free energy 
shown in eq.(\ref{expand}),
we also have to expand the functions $Z_n$ in fractional powers of the
fugacity,
\[Z_1 = \tilde{\lambda} z_1^{(1)} + 
\tilde{\lambda}^{3/2} z_1^{(3/2)} + {\cal O}(\tilde{\lambda}^2),\]
\[Z_2 = z_2^{(0)} + \tilde{\lambda}^{1/2} z_2^{(1)} + 
{\cal O}(\tilde{\lambda}).\]
Introducing the energy scale $\epsilon = \ell_B/a$ with
$a$ being the hard-core radius (or diameter of the spherical particles),
we obtain from eq.(\ref{Z1res})
\begin{equation}
z_1^{(1)} = \frac{\Delta v_0^2}{8 \tilde{\lambda} } = \pi  \epsilon^3 
\end{equation}
\begin{equation}
z_1^{(3/2)} = \frac{\Delta v_0^3}{48 \tilde{\lambda}^{3/2}  } = 
\frac{ (2 \pi  \epsilon^3 )^{3/2}}{6}
\end{equation}
From eq.(\ref{Z2res}) we obtain
\begin{equation}
z^{(0)}_2 =  
\frac{\pi }{3}  \left(2 \epsilon^3 \ {\rm Shi}[\epsilon] 
- {\rm e}^{\epsilon} \left[2+\epsilon+\epsilon^2 \right]
- {\rm e}^{-\epsilon} \left[2-\epsilon+\epsilon^2 \right]
\right) 
\end{equation}
where ${\rm Shi}$ is the hyperbolic sine-integral function defined by
\begin{equation}
{\rm Shi}(z) = \int_0^z {\sinh t \over t} dt
\end{equation}
and
\begin{equation}
z^{(1/2)}_2 =  
\frac{ (2 \pi \epsilon)^{3/2} }{3} 
\left(2 \epsilon^3 ( \Gamma[\epsilon] +2 \gamma + 
\frac{1}{2} \ln[128 \pi \epsilon^3 \tilde{c}] )
- 2 {\rm e}^{-\epsilon} \left[2-\epsilon+\epsilon^2 \right] -
\frac{13}{6} \epsilon^3 \right)
\end{equation}
The coefficient of the $\tilde{\lambda}^{3/2}$ term
in eq.(\ref{expand})  follows from Eq.(\ref{DHres}) as
\begin{equation}
b_{3/2} = - \frac{4}{3} \sqrt{2 \pi}  \epsilon^{3/2},
\end{equation}
the other coefficients are
\begin{equation}
b_2 = - 2 z_1^{(1)} - 2 z_2^{(0)}, 
\end{equation}
\begin{equation}
b_{5/2}  = - 2 z_1^{(3/2)} - 2 z_2^{(1/2)}. 
\end{equation}
In the limit $\epsilon \rightarrow 0$, which corresponds to the 
pure hard-core case, we obtain $ b_2 = +8 \pi /3$, the well-known 
second-virial result. In the strong-coupling limit, $\epsilon \rightarrow 
\infty$, we obtain $b_2 = - 4 \pi  {\rm e}^\epsilon / \epsilon$.
It is seen that the coefficient $b_2$ diverges exponentially as
$\epsilon$ increases.
We attribute this divergence to the formation of
strongly bound ion pairs:
the Boltzmann weight of a pair of 
oppositely charged  ions which stick together is 
$e^\epsilon$, which dominates the statistics for
large values of $\epsilon$.
The next-leading coefficient $b_{5/2}$ is dominated by its
logarithmic contribution and is given by 
$b_{5/2} = -2  (2 \pi)^{3/2} \epsilon^{9/2} \ln( \epsilon)$ 
both in the limit of small and large $\epsilon$. 
In Fig.1 we plot the free energy coefficients $d_2$ and
$d_{5/2}$ as a function of $\epsilon$. For moderately large
values of $\epsilon$, it is seen that the coefficients increase 
exponentially and thus 
render the low-density expansion useless for large  
couplings and large densities. 
It follows that the critical point of the TCP, which is supposed
to occur at coupling strengths of the order of $\epsilon \approx  15$
and a rescaled ion pair density of $\tilde{c} \approx 0.03$\cite{Fisher},
is inaccessible by such a low-density expansion. 

As we will demonstrate in the following section, the OCP does not
contain ion pairing and thus the expansion of the free energy in 
powers of the density is well-behaved even in the limit 
$\epsilon \rightarrow \infty$.

\section{One-component plasma}

We now consider a gas of $N $ identical charged particles, 
interacting with Coulomb  and hard core potentials.
We assume the existence of a uniform oppositely charged
background, which exactly neutralizes the charges. At first,
we will neglect this background charge, and identify and
omit the (infinite) energy contribution of this background in 
our final free energy expression.
The canonical partition function of the system reads:
\begin{equation}
Z_{N}  =\frac{1}{N!} \prod_{i=1}^{N} \left[
\int {\rm d} {\bf r}_i \right] 
\exp \left\{ -\frac 1 2 \int {\rm d} {\bf r}  {\rm d} {\bf r}'
\rho ({\bf r})[w({\bf r-r}') + v({\bf r-r}')]\rho ({\bf r}')
+N [w(0)+v(0)]/2 \right\} 
\label{partition2}
\end{equation}
where $\rho (r)=\sum_{i=1}^{N}\delta (r-r_i) $
denotes the particle density.
The infinite particle self-energies are substracted.
As we did for the TCP,
we perform a double Stratanovitch-Hubbard transform and get
\begin{eqnarray}
Z_{N} &=& \int \frac{{\cal D} \phi}{Z_v} 
\int \frac{{\cal D} \psi}{Z_w} 
\exp \left\{ -\frac 1 2 \int {\rm d} {\bf r}  {\rm d} {\bf r}'
\left[ \psi ({\bf r})w^{-1} ({\bf r-r}')\psi ({\bf r}')+
  \phi({\bf r})v^{-1}({\bf r-r}')\phi({\bf r}') \right] \right\}
  \nonumber \\ && 
\frac{1}{N!} \left[ \int {\rm d} {\bf r}  {\rm e}^{-i \psi({\bf r}) 
-i \phi({\bf r}) +w(0)/2+v(0)/2} \right]^{N}.
\end{eqnarray}
The grand canonical partition function is defined by
\begin{equation}
\label{granddef2}
Z_\lambda = \sum_{N} \lambda^{N} Z_{N} 
\end{equation}
and reads 
\begin{eqnarray}
Z_\lambda &=& \int \frac{{\cal D} \phi}{Z_v} 
\int \frac{{\cal D} \psi}{Z_w} 
\exp \left\{ -\frac 1 2 \int {\rm d} {\bf r}  {\rm d} {\bf r}'
\left[ \psi ({\bf r})w^{-1} ({\bf r-r}')\psi ({\bf r}')+
  \phi({\bf r})v^{-1}({\bf r-r}')\phi({\bf r}') \right] \right.
  \nonumber \\ && \left.
+  \lambda  \int {\rm d} {\bf r} h({\bf r})
{\rm e}^{ v(0)/2 - i \phi({\bf r})}
 \right\}
\end{eqnarray}
with the same definition for $h({\bf r})$ as in eq.(\ref{hdef}).
As for the TCP, the Debye-H\"uckel propagator is defined by its inverse
$v_{\rm DH}({\bf r})= \ell_B {\rm e}^{-\kappa r}/r$, but the screening 
length in this case reads $\kappa^2 = 4 \pi \ell_B \lambda$. The
partition function can be written as
\begin{equation} 
Z_\lambda ={\rm e}^{\lambda V+\lambda V v(0)/2}
\frac{Z_{\rm DH}}{Z_v} \left\langle 
{\rm e}^{  \lambda  \int {\rm d} {\bf r} Q({\bf r})
  }\right\rangle
 \end{equation}
where 
\begin{equation}
Q({\bf r}) \equiv 
h({\bf r}) {\rm e}^{ v(0)/2   - i \phi({\bf r})}
-1-v(0)/2+ \phi^2({\bf r})/2
 \end{equation}
The logarithm of the  partition function
again follows by cumulant expansion,
\begin{equation}
\frac{\ln Z_\lambda}{V} = 2\lambda +\lambda  v (0) + V^{-1} \ln \left[
\frac{Z_{\rm DH}}{Z_v} \right] + \lambda Z_1 +
\frac{ \lambda^2}{2}  Z_2 + \cdots 
\end{equation}
with $Z_1$ and $Z_2$ defined as in eqs.(\ref{Z1def}) and (\ref{Z2def}).
The difference between the Debye-H\"uckel and 
the bare Coulomb self energy reads 
\begin{equation}
\Delta v_0 = v(0)-v_{\rm DH}(0)= 2 \ell_B \sqrt{\pi \ell_B \lambda},
\end{equation}
the functions $Z_1$ and $Z_2$   are given by
\begin{equation}
\label{Z1res2}
Z_1 ={\rm e}^{\Delta v_0/2} -1 - \Delta v_0/2
\end{equation}
\begin{equation}
\label{Z2res2}
Z_2 = \int {\rm d}{\bf r} \left[ 
{\rm e}^{\Delta v_0-w({\bf r})-v_{\rm DH}({\bf r})}
 -{\rm e}^{\Delta v_0}+\frac{1}{2} v^2_{\rm DH}({\bf r})
 \left(1-2 {\rm e}^{\Delta v_0/2}\right) \right]
\end{equation}
The DH contribution to the free energy follows as
\begin{equation}
\label{DHres2}
\lambda  v(0) + V^{-1} \ln \left[
\frac{Z_{\rm DH}}{Z_v} \right] =
\frac{(4 \pi \ell_B \lambda)^{3/2}}{12 \pi}
\end{equation}
Next we have to systematically carry out the back-Legendre
transform to the canonical ensemble. 
To do this, we expand the grand-canonical potential in 
fractional powers of the rescaled fugacity,
\begin{equation}
\frac{a^3 \ln Z_\lambda}{V} = \tilde{\lambda}
 - b_{3/2} \tilde{\lambda}^{3/2} - b_2
\tilde{\lambda}^2 \cdots
\end{equation}
The rescaled concentration of ion pairs, 
$\tilde{c} = a^3 N/V$,
is given by
\begin{equation} \label{Nexpext2}
\tilde{c} = \frac{ \lambda a^3 \partial \ln Z_\lambda/V}
{ \partial \lambda}.
\end{equation}
The fugacity as a function of the concentration
is given by
\begin{equation}
\tilde{\lambda}= \tilde{c}  +\frac{3}{2} b_{3/2} \tilde{c}^{3/2}  +
\left(2 b_2+\frac{27}{8} b_{3/2}^2 \right) \tilde{c}^2 
\cdots
\end{equation}
The free energy density follows as 
\begin{eqnarray}
a^3 f &=& - \frac{ a^3 \ln Z_\lambda}{V}+
\tilde{c} \ln \tilde{\lambda} \nonumber \\
&=& \tilde{c} \ln \tilde{c} -\tilde{c} +d_{3/2} \tilde{c}^{3/2}
 + d_2 \tilde{c}^2 +\cdots
\end{eqnarray}
The prefactors are given by
\begin{equation}
d_{3/2} = b_{3/2} = - \frac{2}{3} \sqrt{\pi}  \epsilon^{3/2},
\end{equation}
\begin{equation}
d_2 = b_2+\frac{9 }{8} b_{3/2}^2 .
\end{equation}

The leading term of $Z_2$ is
\begin{equation}
Z_2   = -\bar{v} + \frac{2 \pi a^3}{3} \left(
\epsilon^3\left[\gamma + \ln \epsilon + \Gamma(\epsilon) 
-\Gamma(3 \sqrt{4 \pi \epsilon \tilde{\lambda}}) -11/6 \right] 
- {\rm e}^{-\epsilon} \left[ 2- \epsilon + \epsilon^2 \right]\right)
+{\cal O}(\tilde{\lambda})
\end{equation}
and the leading term of $Z_1$
\begin{equation}
Z_1   = \frac{\pi \epsilon^3 \tilde{\lambda}}{2} 
+{\cal O}(\tilde{\lambda}^2).
\end{equation}
The infinite constant $\bar{v} = \int {\rm d}{\bf r}v({\bf r})$
corresponds to the interaction of the ions with the oppositely 
charged background and the background self energy and will be neglected 
in the following. The quadratic coefficient of the free energy reads
 \begin{equation}
d_2   = - \frac{ \pi}{3} \left(
\epsilon^3\left[\gamma + \ln \epsilon + \Gamma(\epsilon) 
-\Gamma(3 \sqrt{4 \pi \epsilon \tilde{c} }) -11/6 \right] 
- {\rm e}^{-\epsilon} \left[ 2- \epsilon + \epsilon^2 \right]\right).
\end{equation}
In Fig. 2a we plot this coefficient as a function of $\epsilon$
in the limit of small coupling strengths and setting $\tilde{c} = 1$
inside the logarithm. As one can see,
in the zero-coupling limit, $\epsilon \rightarrow 0$, this
coefficient becomes 
 \begin{equation}
d_2   = \frac{ 2 \pi}{3} ,
\end{equation}
the standard second-virial result. 
We observe a non-monotonic behavior of the coefficient $d_2$,
and for $\epsilon > 3.5$ it becomes in fact negative.

For large values of $\epsilon$, which corresponds to 
strong Coulomb interaction or relatively small hard-core diameter,
 the free energy density
is conveniently  rescaled by the Bjerrum volume $\ell_B^3$ and reads 
\begin{equation}
\ell_B^3 f = \hat{c} \ln \hat{c}  +\hat{d}_{3/2} \hat{c}^{3/2}
 + \hat{d}_2 \hat{c}^2 +\cdots
\end{equation}
where we introduced the rescaled concentration $\hat{c} = \ell_B^3 c$
(which is related  to the so-called plasma parameter $\Gamma$, which is
commonly used to characterize a OCP, by $\Gamma^3 = 4 \pi \hat{c}/3$).
The coefficient appearing in this free energy expansion are
\begin{equation}
\hat{d}_{3/2} =  - \frac{2}{3} \sqrt{\pi},
\end{equation}
\begin{equation}
\hat{d}_2=
- \frac{ \pi}{3} \left(
\gamma + \ln \epsilon + \Gamma(\epsilon) 
-\Gamma(3 \sqrt{4 \pi \hat{c}/\epsilon^2 }) -11/6  -\epsilon^{-3} 
{\rm e}^{-\epsilon} \left[ 2- \epsilon + \epsilon^2 \right]\right).
\end{equation}
In Fig.2b we plot $d_2$ as a function of $\epsilon$ for large
values of $\epsilon$, again setting the concentration appearing
inside the logarithm equal to unity.  In the strong-coupling
limit, $\epsilon \rightarrow \infty$, the coefficient approaches the
value
\begin{equation} \label{Abe}
\hat{d}_2=
- \frac{\pi}{3} \left[ 2 \gamma -11/6 + \ln(3 \sqrt{4 \pi \hat{c}})
\right] ,
 \end{equation}
which agrees exactly with the result obtained by Abe, by
performing an infinite resummation of terms of the Mayer expansion\cite{Abe}.
His calculation was done without hard-cores, and thus constitutes
a special case of our more general result for arbitrary hard-core
diameter. In Fig. 2b we denote the Abe limit, 
$\hat{d}_2 = - \frac{\pi}{3} \left[ 2 \gamma -11/6 + \ln(3 \sqrt{4 \pi})
\right]  \simeq -1.76476$ by a broken line. As one can see, the
convergence towards the asymptotic Abe result is quite slow, and even for
strongly charged spheres with $\epsilon = 10$ the actual coefficient $\hat{d}_2$
has reached only one half of its asymptotic value.

\section{Discussion}

In this paper 
we introduce a novel field-theoretic formulation for plasmas and
charged electrolyte or colloidal solutions.
We are able to  include, in addition to the Coulomb interactions, hard-core
or excluded volume interactions. We apply our method to the
one-component plasma (OCP) and the symmetric two-component plasma
(TCP) and obtain, as a main result, the first few terms in a
systematic low-density expansion of the free energy.
The major advantage over previous calculational methods is that 
each order in the density expansion, which corresponds to an
infinite Mayer sum, is reexpressed as a single diagram involving
a composite, non-linear, but local operator. This entails that
our results are non-perturbative 
with respect to the coupling strength
$\epsilon = \ell_B/a$, and are thus valid both for 
vanishing electric charge, $\epsilon \rightarrow 0$, in
which case we recover the virial expansion for a hard-core gas,
and for vanishing hard-core diameter,  $\epsilon \rightarrow \infty$.
In the strong-coupling limit,  $\epsilon \rightarrow \infty$,
our results for the OCP asymptotically approach the
previous results by Abe, but even for values of the coupling
parameter as large as $\epsilon \simeq 10$ the differences to the
asymptotic result are quite large. The TCP, on the other hand,
does not have a well-defined strong-coupling limit, and we 
find the expansion coefficients of the free energy to increase
progressively as the coupling parameter increases. This fact
is connected with the formation of strongly associated ion pairs.
One of the motivations of the present work was to elucidate the 
critical behavior of the TCP, which shows a critical point at
a coupling strength of roughly $\epsilon \simeq 15$. It is clear
that unless resummation techniques of the low-density expansion are used, 
the low-density expansion cannot be used to predict the
location of this critical point.

\begin{figure}
\caption{
Plot of the coefficients $d_2$ and $d_{5/2}$ in a
systematic and exact low-density expansion of the
free energy for the symmetric two-component plasma
(TCP). The coefficients are shown 
 as a function of the coupling strength $\epsilon =
\ell_B /a$, where $\ell_B = q^2/(4 \pi \varepsilon k_B T)$
is the distance at which two
charged particles interact with thermal energy, 
and $a$ is the hard-core diameter. It is seen that the 
coefficients diverge exponentially as $\epsilon $
increases.}
\caption{ Plot of the quadratic free energy coefficient 
in the small-coupling limit, a), and in the strong coupling 
limit, b). For $\epsilon \rightarrow 0$, one recovers
the hard-core virial coefficient, $d_2 = 2 \pi /3$,
and for $\epsilon \rightarrow \infty$ the Abe result is 
obtained, which is denoted by a broken line.}
\end{figure}


\begin{references}

\bibitem{rev1}
M. Baus and J.-P. Hansen, Phys. Rep. {\bf 59}, 1 (1980);
P. Minnhagen, Rev. Mod. Phys. {\bf 59}, 1001 (1987).

\bibitem{rev2}
C. Deutsch, Y. Furutani, and M.M. Combert,
Phys. Rep. {\bf 69}, 85 (1981);
S. Ichimaru, Rev. Mod. Phys. {\bf 54}, 1017 (1982).

\bibitem{Borukhov}
I. Borukhov, D. Andelman, and H. Orland, Phys. Rev. Lett. 
{\bf 79}, 435 (1997).

\bibitem{Abe}
R. Abe, Prog. Theor. Phys. {\bf 22}, 213 (1959);
E.G.D. Cohen and T.J. Murphy, Phys. Fluids {\bf 12}, 1404 (1969).

\bibitem{OCPhardcore}
E. Waisman and J.L. Lebowitz, J. Chem. Phys. {\bf 56}, 3086 (1972);
F. Lado, Mol. Phys. {\bf 31}, 1117 (1976);
J.-P. Hansen and J.J. Weis, Mol. Phys. {\bf 33}, 1379 (1977);
D. MacGowan, J. Phys. C {\bf 16}, 59 (1983).

\bibitem{Debye}
P.W. Debye and E. H\"uckel, Z. Phys. {\bf 24}, 185 (1923).

\bibitem{Mayer}
J.E. Mayer, J. Chem. Phys. {\bf 18}, 1426 (1950).

\bibitem{Edwards}
S.F. Edwards, Philos. Mag. {\bf 4}, 1171 (1959).

\bibitem{Stell}
G.R. Stell, K.C. Wu, and B. Larsen, Phys. Rev. Lett. {\bf 37},
1369 (1976).

\bibitem{Fisher}
M.E. Fisher and Y. Levin, Phys. Rev. Lett. {\bf 71}, 3826 (1993).

\bibitem{Sengers}
J.M.H. Levelt Sengers and J.A. Given, Mol. Phys. {\bf 80}, 899 (1993).

\bibitem{Brilliantov}
N.V. Brilliantov, C. Bagnuls, and C. Bervillier,
Phys. Lett. A {\bf 245}, 274 (1998).

\bibitem{Netz}
R.R. Netz and H. Orland, Europhys. Lett., in press.


\end{references}
\end{document}